\documentclass[aps,prl,twocolumn,superscriptaddress,longbibliography,amsmath,amssymb]{revtex4-2}

\usepackage{xcolor}
\usepackage{graphicx}
\usepackage{soul}

\newcommand{\suppinfo}{Supplemental Material~\cite{supp-info} }
\begin{document}

\title{Distinguishing different stackings in layered materials via luminescence spectroscopy}
\author{Matteo Zanfrognini}
\affiliation{Dipartimento di Scienze Fisiche, Informatiche e Matematiche, Universit$\grave{a}$ di Modena e Reggio Emilia, I-41125 Modena, Italy}
\affiliation{Centro S3, CNR-Istituto Nanoscienze, I-41125 Modena, Italy}
\author{Alexandre Plaud}
\affiliation{Université Paris-Saclay, ONERA, CNRS, Laboratoire d'étude des microstructures, 92322, Châtillon, France.}
\affiliation{Université Paris-Saclay, UVSQ, CNRS,  GEMaC, 78000, Versailles, France}
\author{Ingrid Stenger}
\affiliation{Université Paris-Saclay, UVSQ, CNRS,  GEMaC, 78000, Versailles, France}
\author{Fr\'ed\'eric Fossard}
\affiliation{Université Paris-Saclay, ONERA, CNRS, Laboratoire d'étude des microstructures, 92322, Châtillon, France.}
\author{Lorenzo Sponza}
\affiliation{Université Paris-Saclay, ONERA, CNRS, Laboratoire d'étude des microstructures, 92322, Châtillon, France.}
\author{L\'eonard Schu\'e}
\affiliation {Laboratoire de Physique de la Matière Condensée, Ecole Polytechnique, CNRS, Institut Polytechnique de Paris, 91120 Palaiseau, France}
\affiliation{Université Paris-Saclay, ONERA, CNRS, Laboratoire d'étude des microstructures, 92322, Châtillon, France.}
\affiliation{Université Paris-Saclay, UVSQ, CNRS,  GEMaC, 78000, Versailles, France}
\author{Fulvio Paleari}
\email{Corresponding author: fulvio.paleari@nano.cnr.it}
\affiliation{Centro S3, CNR-Istituto Nanoscienze, I-41125 Modena, Italy}
\author{Elisa Molinari}
\affiliation{Dipartimento di Scienze Fisiche, Informatiche e Matematiche, Universit$\grave{a}$ di Modena e Reggio Emilia, I-41125 Modena, Italy}
\affiliation{Centro S3, CNR-Istituto Nanoscienze, I-41125 Modena, Italy}
\author{Daniele Varsano}
\affiliation{Centro S3, CNR-Istituto Nanoscienze, I-41125 Modena, Italy}
\author{Ludger Wirtz}
\affiliation{Department of Physics and Materials Science, University of Luxembourg, 1511 Luxembourg, Luxembourg}
\author{Fran\c{c}ois Ducastelle}
\affiliation{Université Paris-Saclay, ONERA, CNRS, Laboratoire d'étude des microstructures, 92322, Châtillon, France.}
\author{Annick Loiseau}
\email{Corresponding author: annick.loiseau@onera.fr}
\affiliation{Université Paris-Saclay, ONERA, CNRS, Laboratoire d'étude des microstructures, 92322, Châtillon, France.}
\author{Julien Barjon}
\email{Corresponding author: julien.barjon@uvsq.fr}
\affiliation{Université Paris-Saclay, UVSQ, CNRS,  GEMaC, 78000, Versailles, France}

\begin{abstract}
Despite its simple crystal structure, layered boron nitride features a surprisingly complex variety of phonon-assisted luminescence peaks. We present a combined experimental and theoretical study on ultraviolet-light emission in hexagonal and rhombohedral bulk boron nitride crystals. Emission spectra of high-quality samples are measured via cathodoluminescence spectroscopy, displaying characteristic differences between the two polytypes. These differences are explained using a fully first-principles computational technique that takes into account
radiative emission from ``indirect'', finite-momentum, excitons via coupling to finite-momentum phonons.
We show that the differences in peak positions, number of peaks and relative intensities can be qualitatively and quantitatively explained, once a full integration over all relevant momenta of excitons and phonons is performed. 
\end{abstract}

\date{\today}
\maketitle

\begin{figure}
\centering
\includegraphics[width=0.45\textwidth]{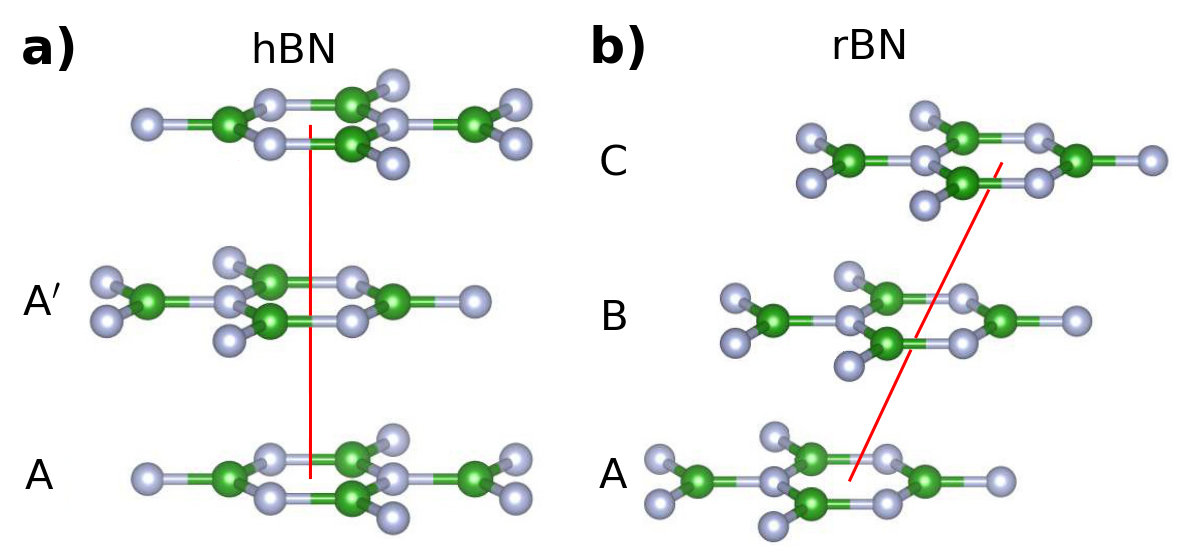}
\caption{Stacking sequences of sp$_2$ BN considered in this work: in a) Boron Nitride with AA$'$ stacking is shown, while in b) the three shifted layers forming rBN unit cell are presented. Nitrogen and Boron atoms are shown in gray and green,respectively.}
\label{stack}
\end{figure}

%
%
Layered boron nitride (BN) crystals are identified as strategic materials for the integration of graphene and 2D semiconductors in optoelectronic devices based on van der Waals heterostructures~\cite{Novoselov2016,Geim2013,Robert2017}.
To this end, largely scalable crystal growth methods able to produce high quality samples are desirable.
The highest quality BN single crystals are mostly grown from a catalytic melt either at high pressure and high temperature (HPHT)~\cite{Solozhenko1996,Watanabe2004,Taniguchi2007} or, more recently, at intermediate or atmospheric pressure and high temperature~\cite{Kubota2007,Liu2018,Onodera2020,Sonntag2020,Maestre2022May}. 
The resulting crystals are limited in size or polycrystalline, which restricts their possible applications to optoelectronics. Up-scalable fabrication techniques at low pressure, such as chemical vapour deposition (CVD) or molecular beam epitaxy (MBE) allow instead for the controlled synthesis of BN thin films on large surfaces. 
However, they have encountered a limited success up to now due to the polymorphism of boron nitride. 
The layered bulk crystal can come, in principle, in six different polytypes~\cite{gil2022polytypes}, with the two most stable ones adopting the hexagonal (hBN) and rhombohedral (rBN) Bravais lattices.
In hBN, two adjacent BN single layers differ by a $\pi$ rotation, resulting in the so-called AA' stacking sequence, where boron and nitrogen atoms sit on top of each other (Figure \ref{stack}a). Conversely, the unit cell of rBN crystals is composed by three BN monolayers, which are rigidly shifted along the same direction by the B-N planar interatomic distance: this stacking motif (ABC sequence) is shown in Figure \ref{stack}b.
While this stacking difference entails an extremely high energy cost associated to the transformation from rBN to hBN~\cite{Yu2003}, these two polytypes are difficult to distinguish experimentally from a crystallographic point of view. 
Even from a computational point of view, the calculated stability difference of the two polytypes is close to the limit of occuracy of modern \textit{ab initio} methods ~\cite{Luo2017,Pedersen2019,gil2022polytypes}.
In addition, the interaction with the substrate affects the abundance of stable rBN and hBN phases in synthetic products~\cite{Sato1985,Sutter2013,Henry2016,Souqui2021}.
For these reasons, the stacking sequence is rarely characterized in recent reports about BN multilayer growth, so that possible differences in the respective optoelectronic properties of the two polytypes might have been overlooked.
\begin{figure*}
\centering
\includegraphics[width=0.8\textwidth]{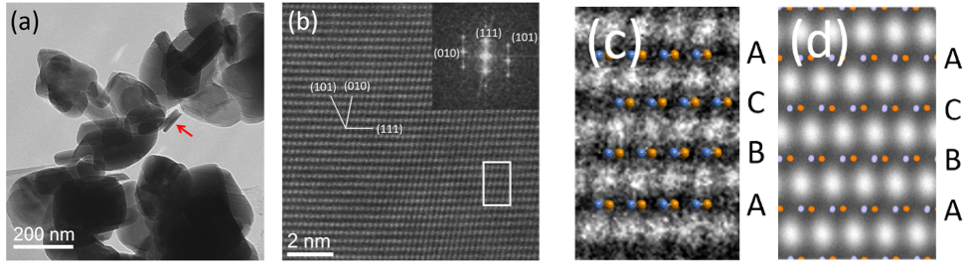}
\caption{(a) Bright field TEM image of the reference rBN powder. (b) High resolution TEM image in the $[10\bar{1}]$ zone axis of the crystallite indicated by the red arrow in (a). The traces of (101), (001) and (111) rBN planes reported with white lines are identified with the Fourier transform plotted in inset. (c) Magnified image of the white rectangle in (b) where the atomic positions of B and N (colored spheres) are deduced from the simulation (d) which has been performed with the illumination conditions used experimentally. Crystallographic notations refer here to the rhombohedral phase (see \suppinfo for details). }
\label{tem}
\end{figure*}
\\In this work, we present a spectroscopic investigation of rBN using cathodoluminescence (CL) spectroscopy. By comparing CL spectra obtained for rBN with analogous results for hBN~\cite{Cassabois2016,Schue2019}, we demonstrate that the stacking sequence affects the emission fine structure of rBN and hBN crystals, making CL an ideal experimental probe to discriminate between the two polytypes. Our experimental observations are explained by \textit{ab initio} calculations of luminescence spectra for the two polytypes, explicitly including exciton-phonon interactions.

\begin{figure}[ht]
\centering
\includegraphics[width=0.48\textwidth]{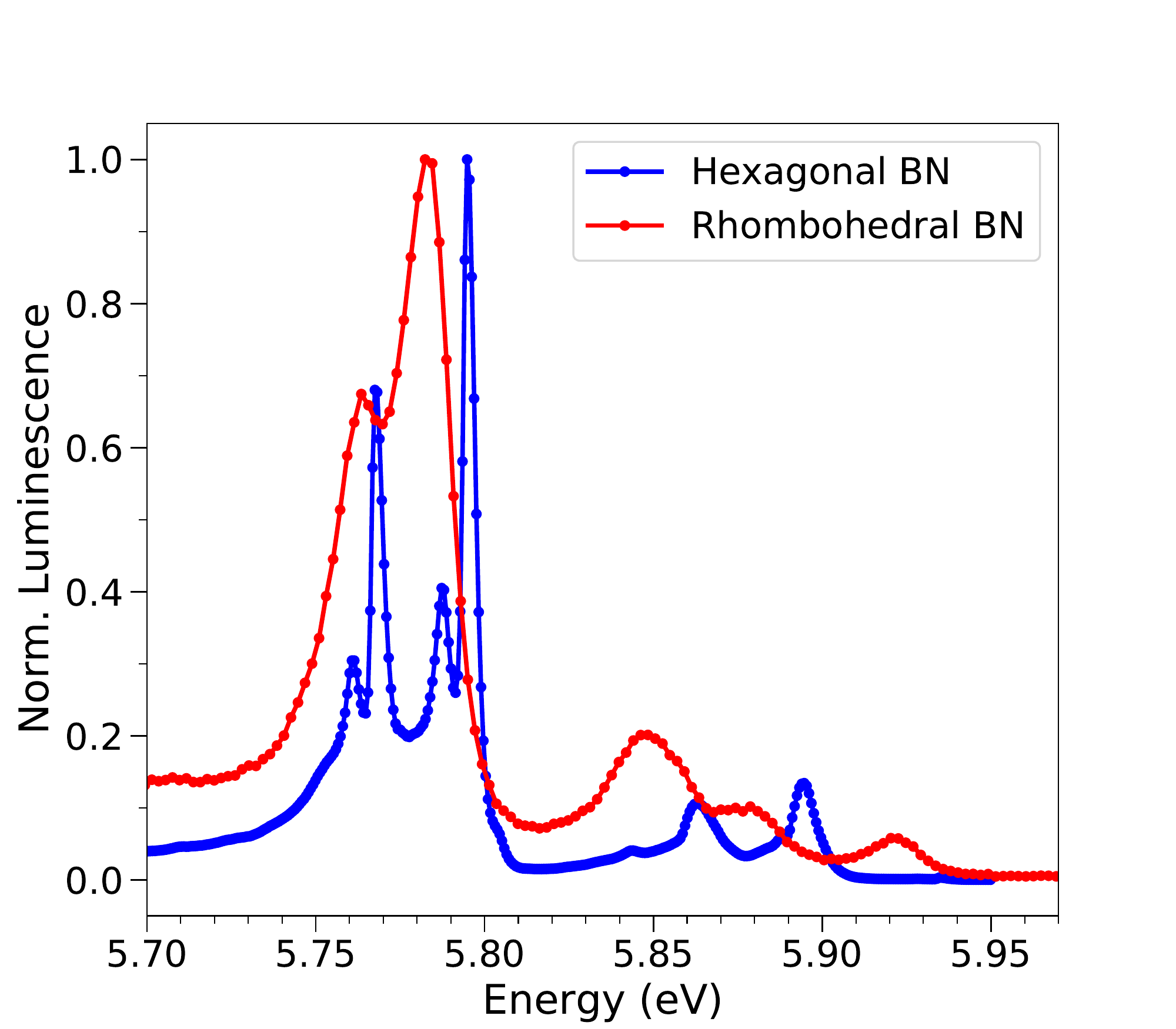}
\caption{Comparison of experimental hBN (blue) and rBN (red) CL spectra at $T=5$ K.}
\label{fig:expCL}
\end{figure}

%
%
The reference sample investigated here is the rBN powder fabricated by T. Sato~\cite{Sato1985}, which is known as the international standard for the crystallographic diffraction database.\footnote{Sample (f) in Ref.~[\onlinecite{Sato1985}] is known under No 00-045-1171 for the Joint Committee on Powder Diffraction Standards (JCPDS) http://www.icdd.com} To our knowledge, this is the highest quality  rBN single crystal available today. Fig.\  \ref{tem} (a) presents a transmission electron microscopy (TEM) image of the powder. It consists of cylindrical rBN crystallites with a typical 200 nm diameter and a ~50 nm thickness. The ABC stacking sequence can be observed in the high-resolution image of the transverse section reported in Fig.\  \ref{tem} (b). The distance between B and N in this projection is 0.072 nm, which cannot be resolved due to our 0.12 nm TEM limit resolution. Nevertheless, the positions of B and N atomic columns can be identified in Fig.\  \ref{tem} (c) thanks to simulations performed in the conditions of the image acquisition in Fig.\  \ref{tem} (d) (see \suppinfo for details). The identification of the rBN structure is further confirmed by comparing its Raman spectrum with the one of hBN as presented in \suppinfo, section Raman spectroscopy. In the following, the properties of the reference rBN sample (ABC stacking) will be compared with a reference hBN crystal grown by HPHT~\cite{Taniguchi2007}. 
\\We now turn to the discussion of the exciton-dominated luminescence spectra as studied by CL using the set up detailed  in \suppinfo.
A comparison between the experimental CL spectra of hBN and rBN at $T=5$ K is shown in Fig. 2. The visible features are due to phonon-assisted excitonic recombinations as will be discussed below.
The two spectra display several key differences, including a redshift of the rBN features with respect to the corresponding hBN ones (which amounts to $15$ meV for the highest peak), and, most importantly, the presence of two relevant structures at 5.847 and 5.919 eV only in rBN. 
The high accuracy of the experimental rBN spectrum is crucial to clearly resolve the fine structure of the intrinsic phonon-assisted peaks~\cite{Moret2021,Gil2022}, enabling us to explain these points in conjunction with the theoretical modelling in the following.
Experimentally, these reported differences are fully significant, as we obtained almost identical spectra from a rBN sample grown by CVD on 6H-SiC. A detailed comparison between the two samples is included in \suppinfo, along with a discussion of the defect peaks appearing in the CL signal measured at frequencies lower than those shown in Fig. \ref{fig:expCL}.
%

\textit{Ab initio} calculations~\cite{Sponza2018} indicate that rBN is an indirect-bandgap insulator. The exciton dispersion resulting from the solution of the Bethe-Salpeter equation (BSE) at finite momentum is indirect as well, its minimum being located near the point $\Omega = [\frac{1}{6},\frac{1}{6},0]$ in the middle of the $\Gamma$K symmetry direction in the hexagonal Brillouin zone (hBZ). According to our calculation, the energy difference between the lowest-lying exciton (due to indirect electronic transition) and the optically accessible (i.e., direct and dipole-allowed) $\Gamma$ excitons is $230$ meV (see \suppinfo for the exciton dispersion curve computed along this direction). This means that the excitonic radiative recombination in rBN requires the assistance of phonons with a wave-vector around the $\Omega$ point, similarly to what happens in hBN.

\begin{figure*}[!t]
  \centering
    \includegraphics[width=0.8\textwidth]{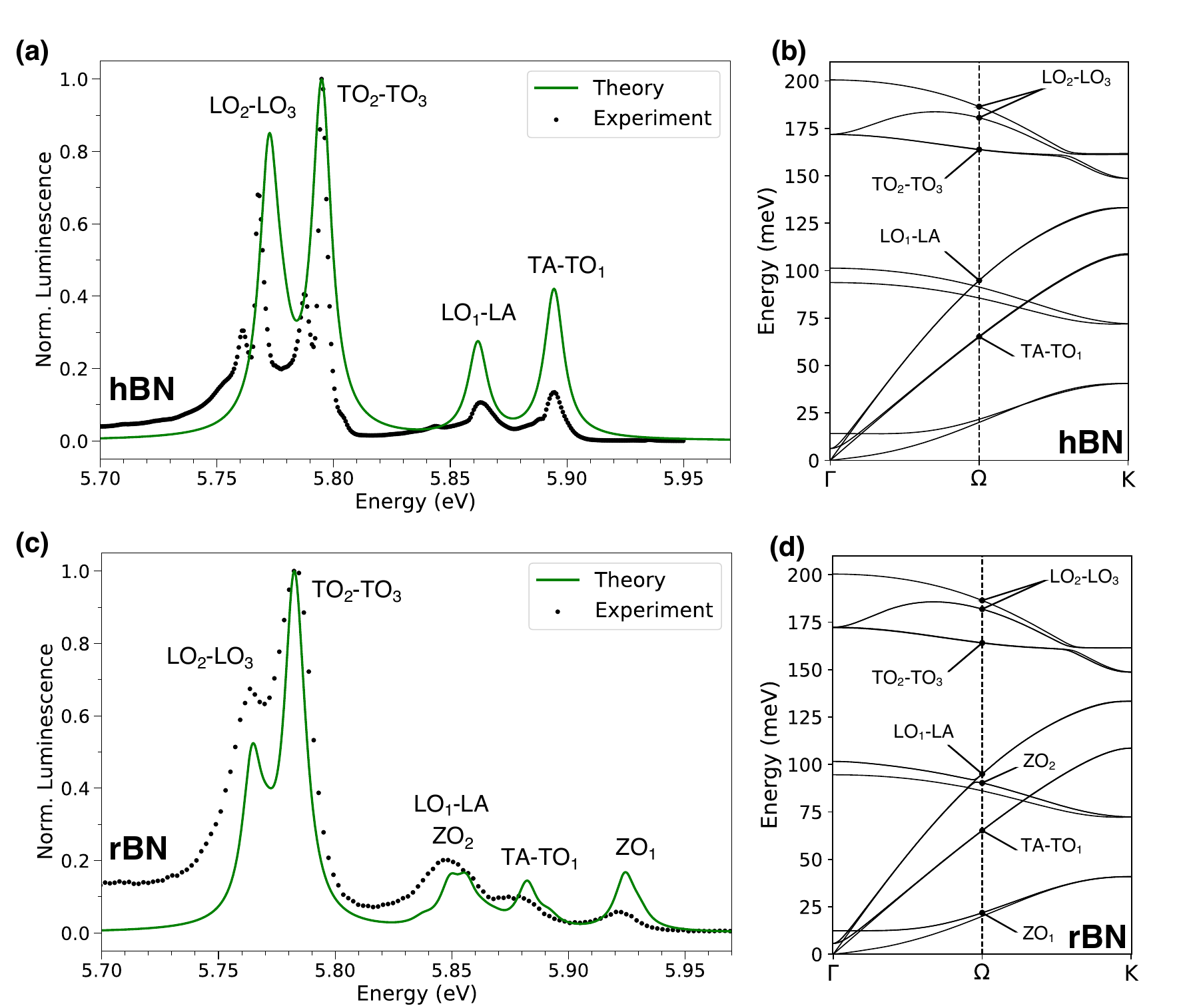}
    \caption{Experimental (black dots) and theoretical (green lines) luminescence spectra for hBN (a) and rBN (c). In both (a) and (c), theoretical spectra are blueshifted by 1.04 eV to match the position of the highest intensity peak in the experimental spectrum. Phonon dispersions in hBN (b) and rBN (d) along the $\Gamma$-K direction: phonon branches contributing to the luminescence spectra are highlighted at the $\Omega$ point, in the middle of the $\Gamma$-K direction. See the main text for the phonon mode labelling. Almost-degenerate phonon branches are paired with a hyphen.}
    \label{fig:Result_LM_hBN_rBN}
\end{figure*}

The theoretical luminescence spectra have been computed using the expression~\cite{ThesisFulvio,Chen2019}:
\begin{equation}
    I(E) \propto \sum_{\lambda}\sum_{\textbf{Q}}\sum_{\nu}N(E_{\lambda}(\mathbf{Q}))\Gamma_{\lambda}^{\nu,\mathbf{Q}}(E)
    \label{eq:EQ_I_1}
\end{equation}
where $\lambda$ is an index running over exciton bands, $\mathbf{Q}$ is the exciton momentum and $\nu$ denotes the phonon branches. $N(E_{\lambda}(\textbf{Q})) = e^{-\frac{E_{\lambda}(\textbf{Q}) - \mu}{k_{B}T_{\mathrm{exc}}}}$ is a Boltzmann distribution representing the exciton population from which light emission occurs, $\mu$ being the energy of the lowest-energy exciton in the system and $T_{\mathrm{exc}}$ the effective excitonic temperature. We fixed $T_{\mathrm{exc}}$ to $20$ K, which is its experimental value obtained for low sample temperatures (below $10$ K, as in Fig. \ref{fig:expCL}). (We have checked that our results are stable with respect to small changes of this parameter.)
\\The probability $\Gamma_{\lambda}^{\nu,\textbf{Q}}(E)$ describes photon emission by a finite-momentum exciton $|\lambda,\mathbf{Q}\rangle$, assisted by a phonon $(\nu,\mathbf{Q})$. This quantity has been computed using second-order time-dependent perturbation theory, similarly to Refs.~[\onlinecite{Cannuccia2019,Lechifflart2022}], only considering phonon emission processes~\cite{Grosso2000} (which dominate at small temperature):
\begin{equation}
    \Gamma_{\lambda}^{\nu,\mathbf{Q}}(E) = \left|T_{\lambda}^{\nu,\mathbf{Q}}\right|^{2} \frac{(1 + n_{\nu,\mathbf{Q}})\delta [E - E_{\lambda}(\mathbf{Q}) + \hbar\omega_{\nu,\mathbf{Q}}]}{E_{\lambda}(\mathbf{Q}) - \hbar\omega_{\nu,\mathbf{Q}}},
    \label{eq:Gamma_eq}
\end{equation}
with 
\begin{equation}\label{eq:osc_str}
    T_{\lambda}^{\nu,\mathbf{Q}}=\sum_{\lambda_2} \frac{D_{\lambda_2} G^{\nu}_{\lambda_2,\lambda}(\mathbf{Q},-\mathbf{Q})}{E_{\lambda_2}(\Gamma) + \hbar\omega_{\nu,\mathbf{Q}} - E_{\lambda}(\mathbf{Q})}.
\end{equation}
In Eqs. \eqref{eq:Gamma_eq} and \eqref{eq:osc_str}, the index $\lambda_2$ runs over the excitonic states at the $\Gamma$ point with energy $E_{\lambda_2}(\Gamma)$. The quantity $D_{\lambda_2}$ is the excitonic optical dipole strength averaged over in-plane polarization directions. $n_{\nu,\textbf{Q}}$ corresponds to the Bose-Einstein phonon occupation factor, while $E$ is the energy of the emitted photon; the Dirac delta guarantees energy conservation and has been numerically approximated with a Lorentzian function with FWHM equal to 5 meV in order to match the experimental peaks.
Finally, the exciton-phonon coupling matrix element $G^{\nu}_{\lambda_2,\lambda}(\textbf{Q},-\textbf{Q})$ describes the scattering amplitude for an exciton $|\lambda,\textbf{Q}\rangle$ to states $|\lambda_2,\Gamma\rangle$ while assisted by phonon mode $\nu$~\cite{Chen2019}:
\begin{equation}\label{eq:G_exc_ph}
\begin{split}
     &G^{\nu}_{\lambda_2,\lambda}(\textbf{Q},-\textbf{Q}) = \\ &\sum_{vcc'\textbf{k}} A_{\lambda_2}^{*\Gamma}(v\textbf{k},c\textbf{k})A_{\lambda}^{\textbf{Q}}(v\textbf{k},c'\textbf{k}+\textbf{Q})g_{cc'}^{\nu}(\textbf{k}+\textbf{Q};-\textbf{Q}) \\ 
     -&\sum_{vv'c\textbf{k}}A_{\lambda_2}^{*\Gamma}(v\textbf{k},c\textbf{k})A_{\lambda}^{\textbf{Q}}(v'\textbf{k}-\textbf{Q},c\textbf{k})g_{v'v}^{\nu}(\textbf{k};-\textbf{Q}),
\end{split}
\end{equation}
where $A_{\lambda}^{\textbf{Q}}(v\textbf{k}_h,c\textbf{k}_c)$ is the envelope function for exciton $|\lambda,\textbf{Q}\rangle$, with $v,v'$ $(c,c')$ running over the valence (conduction) states and $\mathbf{k}$ being the electronic wave vector in the hBZ. The electron-phonon coupling matrix element $g^{\nu}_{n,n'}(\textbf{k},\textbf{Q})$ represents the scattering between single-particle states $|n',\textbf{k}\rangle$ and $|n,\textbf{k}+\textbf{Q}\rangle$~\cite{Giustino2017}.
Importantly, within our numerical methodology, $G^{\nu}_{\lambda_2,\lambda}(\textbf{Q},-\textbf{Q})$ is computed using the same single-particle Kohn-Sham states both for electron-phonon and excitonic quantities, thus overcoming phase mismatch problems as described in Ref.~[\onlinecite{Lechifflart2022}].
The $\mathbf{Q}$-integration appearing in Eq. \eqref{eq:EQ_I_1} has been performed in local neighbourhoods of the symmetry-equivalent $\Omega$ points corresponding to the excitonic dispersion minima in the hBZ. The computational details \footnote{The theoretical spectra have been obtained using {\tt {Quantum Espresso}}\cite{Giannozzi2009,Giannozzi2017} and {\tt {Perturbo}}\cite{Zhou2021} packages to evaluate ground state electronic properties,  vibrational excitations and electron-phonon matrix elements while exciton energies and wavefunctions have been obtained using {\tt {Yambo}}~\cite{Marini2009,Sangalli2019} code.} needed to reproduce the theoretical results are provided in the \suppinfo. 

In Figure \ref{fig:Result_LM_hBN_rBN}, we present the comparison between experimental CL spectra (black dots) and theoretical BSE results (continuous green lines) for hBN (Fig. \ref{fig:Result_LM_hBN_rBN}a) and rBN (Fig. \ref{fig:Result_LM_hBN_rBN}c).
Figures \ref{fig:Result_LM_hBN_rBN}b and \ref{fig:Result_LM_hBN_rBN}d show the calculated in-plane phonon dispersion along the $\Gamma$K direction for hBN and rBN, respectively.
We find very good agreement between experimental and theoretical data.
The relative energy shift between the two spectra is reproduced theoretically. As the phonon energies in the two systems differ only for a few meV, the $15$ meV shift closely matches the underlying difference between the lowest-lying, finite-momentum exciton levels (which is around $12$ meV). In turn, this difference can be traced back to the combined effects of rBN having both a smaller quasiparticle band gap (by $166$ meV) and exciton binding energy (by $150$ meV) with respect to hBN around the $\Omega$ points in momentum space.
In both hBN and rBN, the spectra are dominated by the two peaks in the low-energy part of the spectrum.
These are phonon-assisted satellites due to longitudinal optical phonons -- denoted as LO$_2$-LO$_3$ modes in the phonon dispersion -- and transverse optical ones (the almost-degenerate pair~\cite{Molina2011} TO$_2$-TO$_3$). 
For hBN, these assignments are in good agreement with the results obtained in Refs.~[\onlinecite{Fulvio2019,Cannuccia2019}], using a finite-difference approach.
Furthermore, the experimental intensity ratio between these peaks is well-reproduced by \textit{ab initio} calculations, with the LO peak being less intense than the TO one.
The additional overtones appearing in the measurements in this energy region are due to higher-order scattering processes~\cite{Vuong2017} and are thus not captured by our theoretical approach, which is restricted to first-order exciton-phonon interaction. The phonon branches involved in the emission process are explicitly labelled in Figs. \ref{fig:Result_LM_hBN_rBN}b and \ref{fig:Result_LM_hBN_rBN}d for the $\Omega$ point only.\footnote{In our labeling of the phonon modes, we chose to disentangle explicitly the almost-degenerate Davydov pairs of modes. This is the reason why, for example, we consider the lowest-energy phonon mode to be a pair of acoustic (ZA) and optical (ZO$_1$) out-of-plane modes, with only the latter being responsible for the signal in Fig. \ref{fig:Result_LM_hBN_rBN}c. In the experimental literature, this pair is usually labeled jointly as ``ZA'', and the same goes for the other pairs.} 
Luminescence spectra of hBN and rBN are qualitatively different at higher energies, as confirmed by \textit{ab initio} results. 
In the case of hBN, we observe only two main peaks: the first (at about 5.86 eV) corresponds to a replica of the LO$_1$-LA phonons, while the higher intensity structure at 5.89 eV is mainly due to TO phonons, with a small contribution from the almost-degenerate transverse acoustic mode (TA-TO$_1$). \textit{Ab initio} results reproduce with great accuracy both the splitting between these peaks and their intensity ratio (the LO$_1$-LA peak being less pronounced than the TO$_1$-TA one), while they tend to overestimate their relative strengths, with respect to the dominant, low-energy satellites. (The agreement may be further improved with a more complete $\textbf{Q}$-point integration in Eq.~\eqref{eq:EQ_I_1}.) We also note that, in agreement with the group theory analysis discussed in Ref.~[\onlinecite{Fulvio2019}], no contributions from the out-of-plane phonon modes appear in the luminescence spectra. This selection rule, which is strictly respected by Eq. \eqref{eq:G_exc_ph}, can be slightly broken in a real experiment, leading to the appearance of a very small signal corresponding to this mode (usually 100 times smaller than the other peaks~\cite{Vuong2017B}).
\\In the case of rBN, the high-energy portion of the CL spectrum shows three large peaks, respectively at about 5.847 eV, 5.878 eV and 5.919 eV, instead of the two peaks appearing in hBN. They are also recovered in the \textit{ab initio} results. The first structure is a combination of phonon-assisted replicas due to the almost-degenerate LA-LO$_1$ branches, albeit with a relevant contribution from optical out-of-plane modes (denoted as ZO$_2$; see \suppinfo for a mode-resolved spectrum). 
Conversely, the peak at 5.878 eV is associated to the TA-TO$_1$ phonons in analogy with the hBN case. 
We emphasise that \textit{ab initio} results correctly reproduce the intensity ratio among these peaks. 
Finally, the highest-energy structure at 5.919 eV turns out to be due to the out-of-plane optical mode ZO$_1$. This is forbidden for the centrosymmetric hBN luminescence while it is allowed in the rBN case because of the lowered symmetry of the crystal lattice.


In conclusion, we have demonstrated that cathodoluminescence is a viable tool to characterize fundamentally similar BN polytypes, which are hardly distinguishable otherwise.
We have explained both experimentally and theoretically how the radiative emission spectrum is affected by the interaction between electronic excitations and lattice vibrations in rhombohedral and hexagonal boron nitride, two prototypical polytypes of low-dimensional layered materials with indirect band gap.
Using a first-principles methodology which accounts for exciton-phonon interactions beyond the state of the art, we are able to provide a comprehensive and accurate description of the finite-momentum exciton states and phonon modes involved, thus showing the discriminating role of out-of-plane lattice vibrations assisting excitonic radiative recombination for rBN but not for hBN.  
We believe that our analysis and methodology could be useful for the growth and characterization of indirect-gap layered materials, which find widespread application as basic building blocks in novel 2D optoelectronic devices.

\begin{acknowledgments}The authors would like to thank  C. Vilar for the technical support on electron microscopy and K. Watanabe and T. Taniguchi for kindly providing a part of the rBN reference powder of T. Sato, M. Chubarov and A. Henry for providing rBN whiskers on 6H-SiC. 
We thank C. Attaccalite and P. Lechifflart for useful discussions about exciton-phonon coupling calculations. This project has received funding from the European Union Horizon 2020 research and innovation programme under grant agreement No 785219  and No 881603 (Graphene Flagship core 2 and core 3), the French National Agency for Research (ANR) under grant agreement No ANR-14-CE08-0018 (GoBN: Graphene on Boron Nitride Technology), MaX -- MAterials design at the eXascale -- a European Centre of Excellence funded by the European Union's program HORIZON-EUROHPC-JU-2021-COE-01 (Grant No. 101093374). D.V. and M.Z. also acknowledge financial support from ICSC – Centro Nazionale di Ricerca in High Performance Computing, Big Data and Quantum Computing, funded by European Union –
NextGenerationEU – PNRR and the Italian national program PRIN2017 grant n. 2017BZPKSZ. L.W. acknowledges funding by Fond National de Recherche (FNR), Luxembourg via project INTER/19/ANR/13376969/ACCEPT. We acknowledge EuroHPC Joint Undertaking for awarding us access to MeluXina at LuxProvide, Luxembourg and CINECA for computational resources, awarded via the ISCRA Grants.\\ 


A.P. and M.Z. contributed equally to this work.
\end{acknowledgments}

%

\end{document}


\title{Supplemental Materials of:  Distinguishing different stackings in layered materials via luminescence spectroscopy}

\author{Matteo Zanfrognini}
\affiliation{Dipartimento di Scienze Fisiche, Informatiche e Matematiche, Universit$\grave{a}$ di Modena e Reggio Emilia, I-41125 Modena, Italy}
\affiliation{Centro S3, CNR-Istituto Nanoscienze, I-41125 Modena, Italy}
\author{Alexandre Plaud}
\affiliation{Université Paris-Saclay, ONERA, CNRS, Laboratoire d'étude des microstructures, 92322, Châtillon, France.}
\affiliation{Université Paris-Saclay, UVSQ, CNRS,  GEMaC, 78000, Versailles, France}
\author{Ingrid Stenger}
\affiliation{Université Paris-Saclay, UVSQ, CNRS,  GEMaC, 78000, Versailles, France}
\author{Fr\'ed\'eric Fossard}
\affiliation{Université Paris-Saclay, ONERA, CNRS, Laboratoire d'étude des microstructures, 92322, Châtillon, France.}
\author{Lorenzo Sponza}
\affiliation{Université Paris-Saclay, ONERA, CNRS, Laboratoire d'étude des microstructures, 92322, Châtillon, France.}
\author{L\'eonard Schu\'e}
\affiliation {Laboratoire de Physique de la Matière Condensée, Ecole Polytechnique, CNRS, Institut Polytechnique de Paris, 91120 Palaiseau, France}
\affiliation{Université Paris-Saclay, ONERA, CNRS, Laboratoire d'étude des microstructures, 92322, Châtillon, France.}
\affiliation{Université Paris-Saclay, UVSQ, CNRS,  GEMaC, 78000, Versailles, France}
\author{Fulvio Paleari}
\email{Corresponding author: fulvio.paleari@nano.cnr.it}
\affiliation{Centro S3, CNR-Istituto Nanoscienze, I-41125 Modena, Italy}
\author{Elisa Molinari}
\affiliation{Dipartimento di Scienze Fisiche, Informatiche e Matematiche, Universit$\grave{a}$ di Modena e Reggio Emilia, I-41125 Modena, Italy}
\affiliation{Centro S3, CNR-Istituto Nanoscienze, I-41125 Modena, Italy}
\author{Daniele Varsano}
\affiliation{Centro S3, CNR-Istituto Nanoscienze, I-41125 Modena, Italy}
\author{Ludger Wirtz}
\affiliation{Department of Physics and Materials Science, University of Luxembourg, 1511 Luxembourg, Luxembourg}
\author{Fran\c{c}ois Ducastelle}
\affiliation{Université Paris-Saclay, ONERA, CNRS, Laboratoire d'étude des microstructures, 92322, Châtillon, France.}
\author{Annick Loiseau}
\email{Corresponding author: annick.loiseau@onera.fr}
\affiliation{Université Paris-Saclay, ONERA, CNRS, Laboratoire d'étude des microstructures, 92322, Châtillon, France.}
\author{Julien Barjon}
\email{Corresponding author: julien.barjon@uvsq.fr}
\affiliation{Université Paris-Saclay, UVSQ, CNRS,  GEMaC, 78000, Versailles, France}

\date{\today}

\maketitle
\subsection{Cathodoluminescence setup} 
The cathodoluminescence set-up combines an optical detection system (Horiba Jobin Yvon SA) and a JEOL7001F field-emission gun scanning electron microscope (FEG-SEM) equipped with a Faraday cup for the beam current measurement. Samples are mounted on a GATAN cryostat SEM-stage to cool them at temperatures from 300 K down to 5 K thanks to cold-finger cryostat with liquid helium. The CL signal is collected by a parabolic mirror and focused with mirror optics on the open entrance slit of a 55 cm-focal length monochromator. The all-mirror optical set-up provides an achromatic focusing with a high spectral sensitivity down to 190 nm. A nitrogen-cooled charge-coupled detector (CCD) silicon camera is used to record CL spectra. All spectra shown in this work are corrected from the spectral sensitivity of the detection set-up. 

\subsection{Reproducibility of luminescence features associated to rBN excitons}

\begin{figure}
\centering
\includegraphics[width=1\columnwidth]{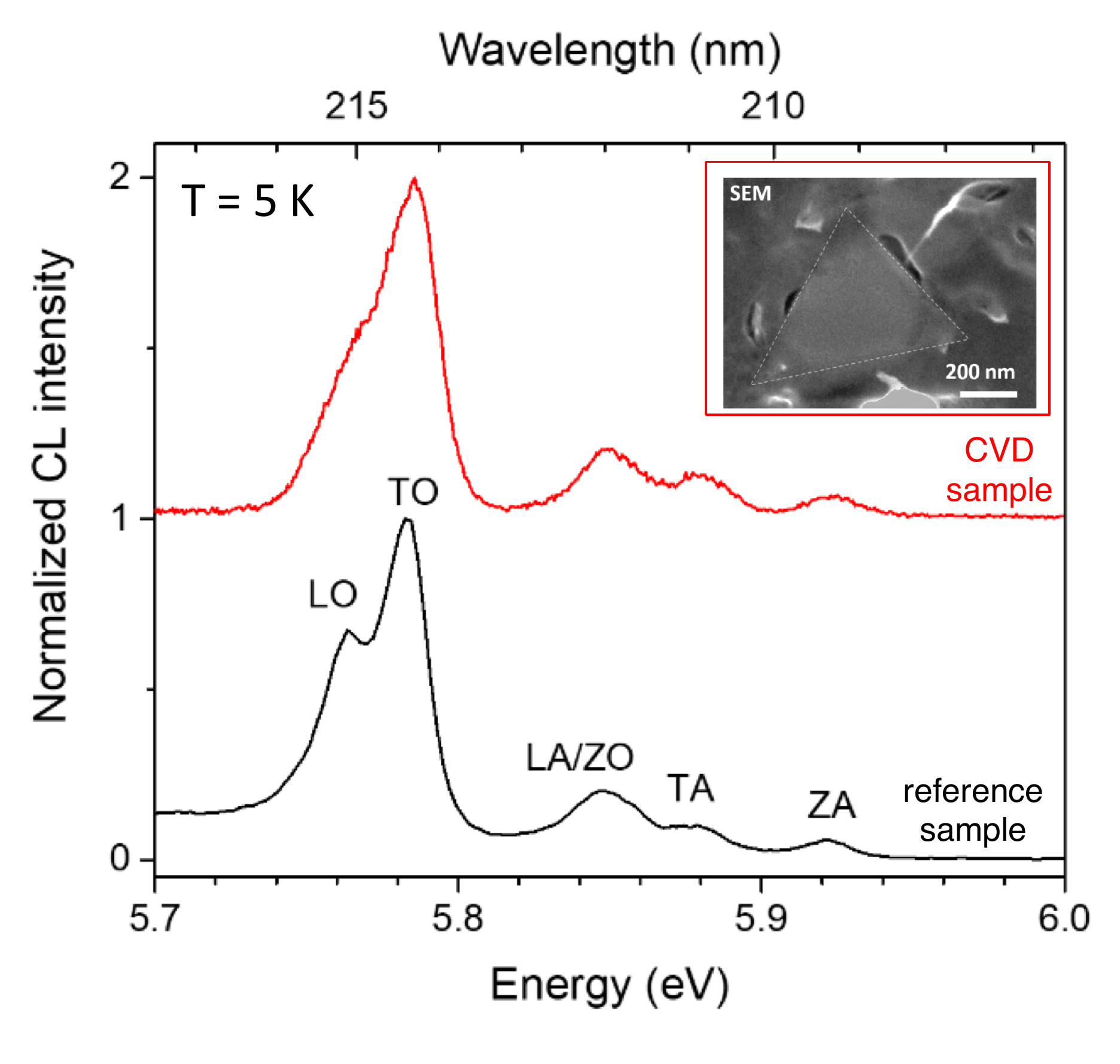}
\caption{Comparison between luminescence from the additional rBN sample considered in this work (red line, Adapted from Ref. [\onlinecite{Schue2017}], SEM image in the inset) and from the reference sample discussed in the main text (black line) at T= 5 K. The voltage and current used were 4 kV and 2 nA, respectively. Note that, at variance with the main text, in this figure we used a pairwise labeling of the almost-degenerate phonon modes. For example, the (ZA,ZO$_1$) pair is collectively labeled ``ZA''.}\label{fig:fig1_SM}
\end{figure}

\begin{figure}
\centering
\includegraphics[width=1\columnwidth]{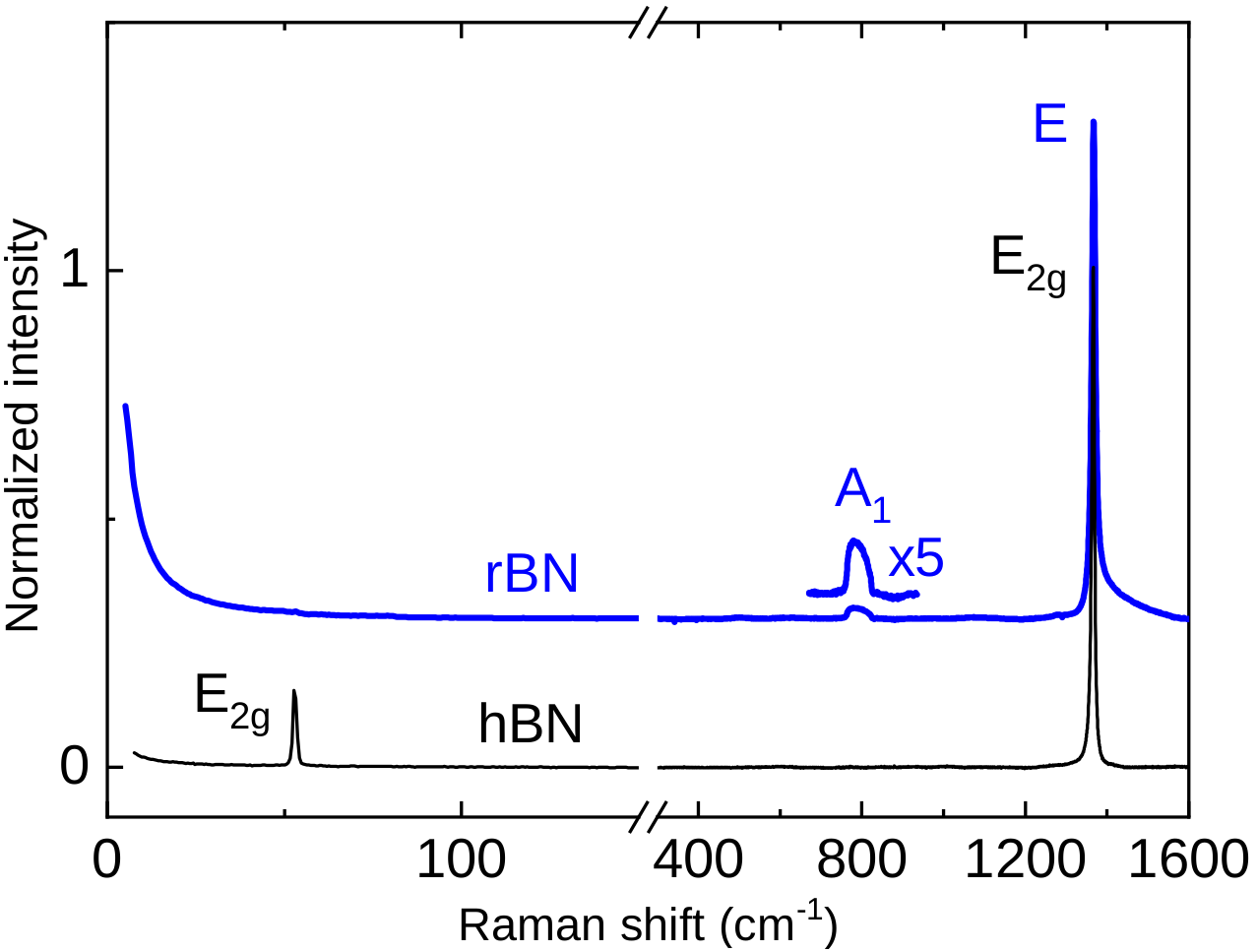}
\caption{Raman diffusion spectra of rBN (blue line) and hBN (black line) samples, where the phonon modes are labelled according to the irreducible representations of point groups C$_{3\mathrm{v}}$ and D$_{6\mathrm{h}}$, respectively. The spectra are normalized to the 1366 cm-1 peak intensity and shifted vertically for clarity.}
\label{fig:fig2_SM_raman}
\end{figure}

For validation purposes, we also performed CL measurements on a second rBN sample. It consists of well-faceted rBN whiskers embedded in an amorphous turbostratic BN matrix, grown by low-pressure CVD on 6H-SiC substrates\cite{Chubarov2015}. 
The CL spectra from both samples are plotted in Fig. \ref{fig:fig1_SM}, the one for the CVD sample, in red, being recorded at the center of the triangular rBN whisker seen in the inset. 
It shows slightly larger linewidths than our reference sample described in the main text. 
This is probably related to bandgap fluctuations due to the lower crystallinity of the second rBN sample. A higher structural disorder is indeed attested, considering the larger linewidth of the Raman peaks observed in rBN once compared to those obtained in hBN.
However, their energies match those reported on the reference sample at the order of the meV. Additionally, the relative intensities between phonon lines are reproduced with less than a 10 \% difference, with respect to the reference sample. This confirms the intrinsic nature of the rBN spectral features investigated in this work.

%
%
%
%
%
%

\begin{figure}
\centering
\includegraphics[width=0.8\columnwidth]{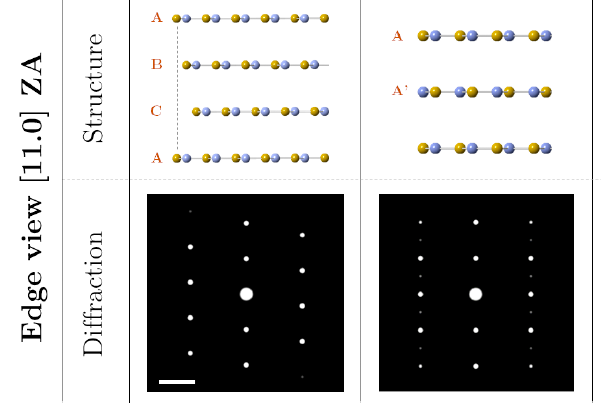}
\caption{Projected atomic structures of rBN (ABC stacking) and hBN (AA' stacking) phases in the $[10\bar{1}]$ zone axis and their corresponding diffraction patterns. Crystallographic notations refer to the hexagonal unit cell.}
\label{fig:fig7_SM_tem}
\end{figure}

\subsection{Raman spectroscopy of rBN}
The Raman spectra of rBN and hBN are compared in Figure \ref{fig:fig2_SM_raman} (blue and black lines, respectively). Both spectra are dominated by an intense peak at the same frequency (1366 cm$^{-1}$), which is used to normalize the spectra.
We can recognize small differences between the Raman signals measured in the two cases. Firstly, in rBN we notice the presence of a weak peak due to the 790 cm$^{-1}$ A$_1$-mode, which is instead inactive in hBN: this result has been already reported for rBN in Ref.~\onlinecite{Liu1995}. Secondly, in agreement with the theoretical predictions of Ref.~\onlinecite{Yu2003}, we notice the absence in rBN of the low frequency Raman peak observed in hBN, due to an E$_{2g}$ low-energy phonon mode.
Note that rather long integration times -- few minutes to several hours depending on the crystal thickness -- are required to detect these low intensity modes since the large band gaps of rBN and hBN do not permit reaching resonant conditions for Raman scattering.
We note that the linewidths of the main peaks in the collected Raman spectra are slightly different in the two crystals: more precisely, we found a linewidth of 10 cm$^{-1}$ for rBN and  only 7 cm$^{-1}$ for hBN. This could suggest a higher structural disorder in rBN than in hBN and in fact the linewidths of the CL spectra in the reference rBN sample appear to be broader than those of hBN grown by HPHT.
%
%
%
\subsection{Transmission Electron Microscopy of rBN}
Transmission Electron Microscopy has been performed on a Zeiss LIBRA 200 TEM. This microscope is equipped with an electrostatic monochromator in the electron gun and an in-column omega filter to achieve energy-filtered images.
The microscope is operated at 200 kV and thanks to the monochromator, the energy dispersion of the electron is reduced to 0.25 eV which reduces the chromatic aberration.
The three-lenses condenser system allows for parallel Köhler illumination on the sample and a 0.12 nm lattice resolution is reached.
The sample is hold in a GATAN rotation-tilt sample holder in order to reach the $[10\bar{1}]$ zone axis.
Images are recorded on an Ultrascan 1000 2048*2048 pixels GATAN camera with a 16-bits dynamic.
Experimental images correspond to a projection of the structure on the $[10\bar{1}]$ zone axis, where the indices refer to the crystallographic notations of the rhombohedral primitive cell. This zone axis provides a view from the side of the honeycomb lattice, revealing the stacking order of the constituent planes. The stacking axis is the $[111]$ axis in the crystallographic notation of the primitive cell. 
These experimental images were analysed with the support of image simulations of the rBN structure in the  $[10\bar{1}]$ zone axis performed with the JEMS software in the Bloch wave formalism and with the experimental illumination conditions used for image acquisition.\cite{Stadelmann1987}
In the HRTEM mode used here, the image is a phase contrast image resulting from the elastic scattering of the electrons due to the atomic potentials. The result is an image of these potentials projected on the  $[10\bar{1}]$ zone axis, convoluted with the instrumental response of the microscope.
With experimental conditions used here, the projected atomic positions indicated with blue and orange bullets in Fig. 2 (c) and Fig. 2 (d) of the main text for B and N respectively correspond to intensity minima (i.e. dark contrast). Nevertheless, their relative positions along the stacking axis is revealed by the position of the intensity maxima. Each edge-on plane is imaged by a line of white dots periodically spaced. From one line to the next one, the dot position is translated by one third of the spacing, indicating a three-layer vertical periodicity. This sequence exactly corresponds to the ABC stacking, specific of the rBN phase. By contrast, if the structure would have been the hBN phase, no shift in the dot position would have been observed from one white dots line to the next one.
Finally, it is worth noticing that the rhombohedral structure can also be described using a hexagonal unitary (but not primitive) cell, facilitating the comparison between rBN, hBN and AB stacking. Because of this, we always performed the simulations of TEM images in the hexagonal cell. In this case, the stacking axis $[111]$ of rBN corresponds to the $c$ axis of the hexagonal cell, and the $[10\bar{1}]$ axis revealing the stacking sequence is the $[11.0]$ axis in the hexagonal reference common to the three structures. Figure \ref{fig:fig7_SM_tem} provides the atomic positions of the rBN and hBN structures in this zone axis together with the corresponding electron diffraction patterns.
\begin{figure}
\centering
\includegraphics[width=0.5\textwidth]{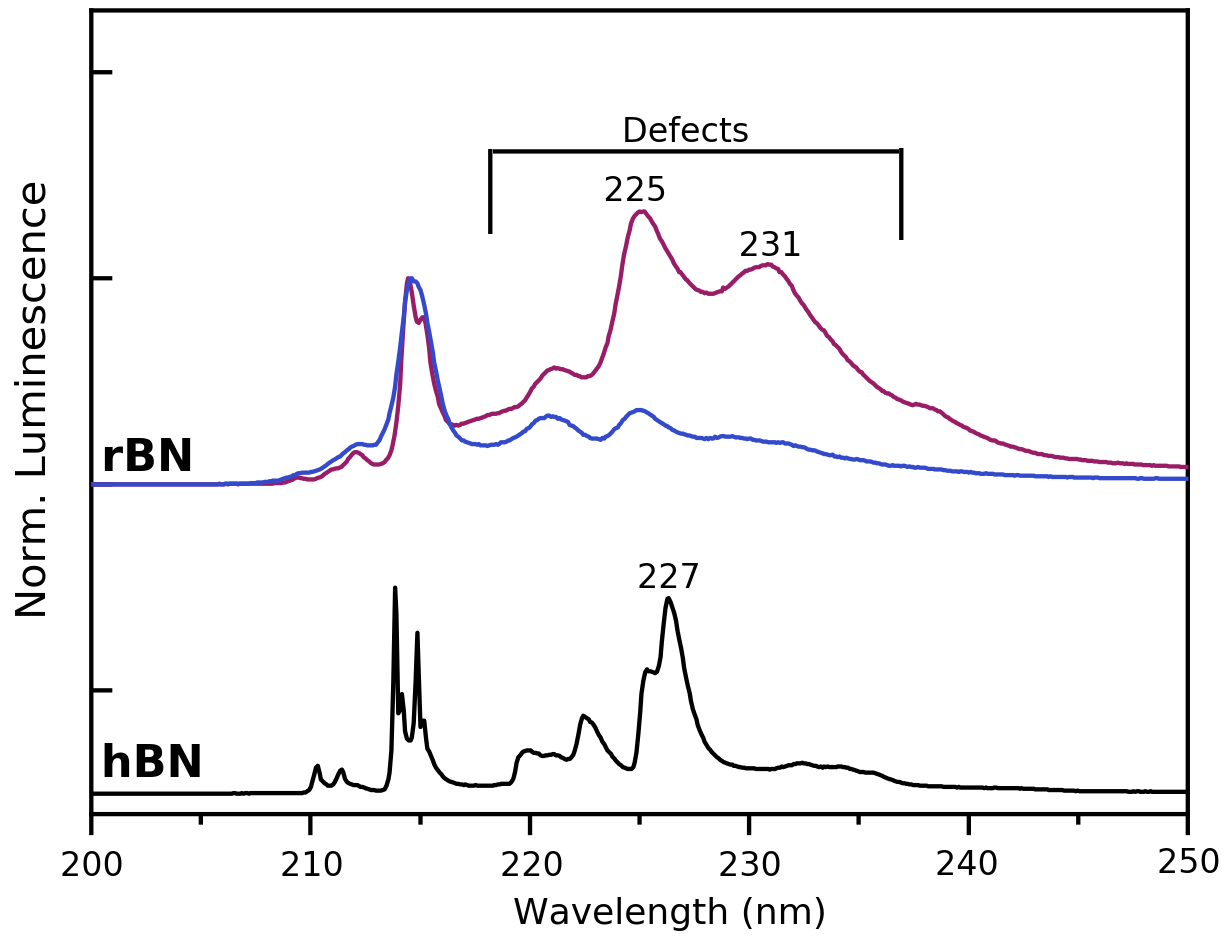}
\caption{Cathodoluminescence signal from rBN and hBN, measured at lower frequencies than those shown in the main text. The structures obtained at lower photon energies are interpreted in both hBN and rBN as due to defect states. In rBN, the shown spectra refer to two different crystallites.}
\label{fig:Lum_defects}
\end{figure}
\subsection{Defect states}
We complement our analysis of luminescence in rBN crystals considering the signal measured at frequencies lower than those discussed in the main text, where emission related to defects and deep levels is usually found. Interestingly, in the low-frequency range we could detect new signals at 225 and 231 nm whose intensities were found to vary among the rBN crystallites, as shown in Fig. \ref{fig:Lum_defects}. In hBN, the corresponding emissions are called the D series with maximum at 227 nm: the D series in hBN is observed in the presence of structural defects such as stacking faults and dislocations \cite{Watanabe2006, Jaffrennou2007}. By analogy, the 225 and 231 nm lines might be associated to the structural defects of rBN: these two lines have been observed in multiwall BN nanotubes \cite{Pierret2015} known to be faceted with a defective rhombohedral stacking \cite{Bourgeois2000} and in epitaxial layers of BN on Ni(111), where the ABC stacking was identified together with ABA faults \cite{Andrieux2020}. A theoretical characterization of these additional emission lines may be achieved by including the treatment of defect states in our numerical approach.~\cite{Libbi2022}

%
%
%
%
%

\begin{figure}[b]
  \centering
    \includegraphics[width=0.8\columnwidth]{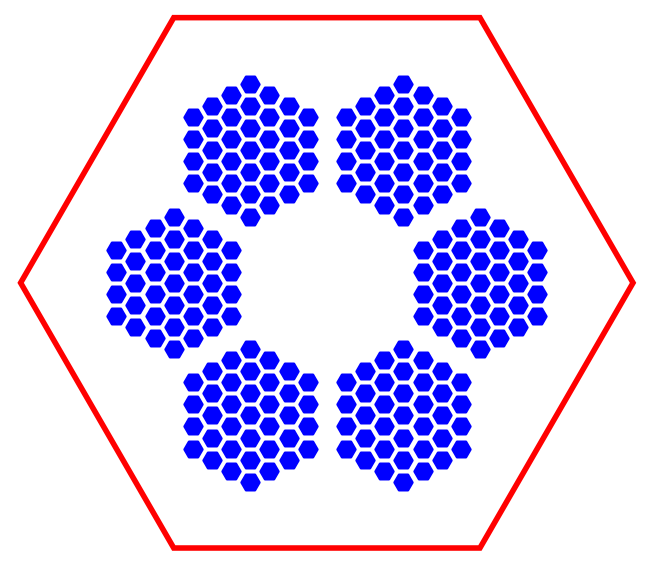}
    \caption{Grid of $\textbf{Q}$-points where finite-momentum excitons have been computed. The contribution to  the luminescence spectra of excitons outside the selected regions is negligible because of the exponentially small occupation factor for the considered value of the effective excitonic temperature T$_{\mathrm{exc}}$.
    \label{fig:Q_grid_LM}
    }
\end{figure}

\begin{figure*}[ht]
  \centering
    \includegraphics[width=1\textwidth]{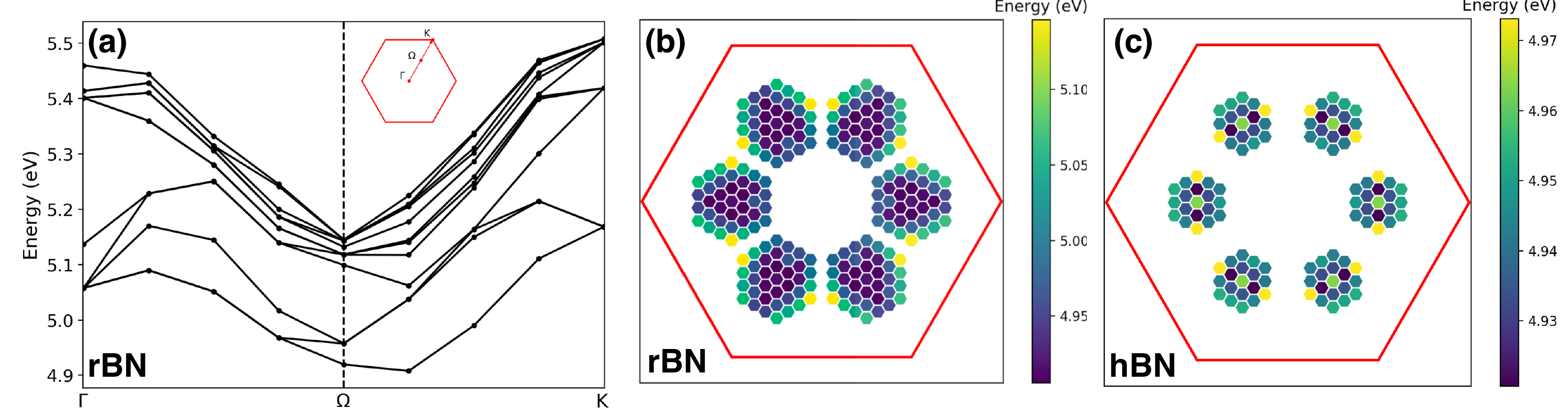}
    \caption{(a) Exciton dispersion along the $\Gamma$K direction in rBN. Black dots represent the computed exciton energies, while lines are guides to the eyes. (b) and (c): values of the exciton energies in the 2D neighborhoods of the $\Omega$ points where the $Q$-summation in Eq. (1) of the main text was conducted.}
    \label{fig:rBN_exc_disp}
\end{figure*}

\subsection{\textit{Ab initio} calculations}

Ground state electronic and structural properties of hBN and rBN have been computed using DFT as implemented in the {\tt{Quantum Espresso}} package~\cite{Giannozzi2009,Giannozzi2017}, using the Local-Density-Approximation (LDA) for the exchange-correlation potential and employing norm-conserving pseudopotentials from the Pseudo-Dojo repository~\cite{VanSetten2018}. Furthermore, in all DFT calculations we have sampled the Brillouin zone with a 12x12x4 Monkhorst-Pack k-grid, using a 100 Ry kinetic energy cutoff for the wavefunctions; lattice parameters and atomic positions have been relaxed up to when forces acting on atoms were smaller than 10$^{-5}$ a.u.
We note that instead of the rhombohedral unit cell, we used an equivalent hexagonal unit cell for rBN (fully containing the three inequivalent atomic layers), because it is analogous to the hBN case and facilitates the analysis and comparison of corresponding $Q$-points between the two systems.
\\Starting from the ground-state charge density, we have computed phonon modes on a 24x24x2 $\textbf{Q}$ grid, using Density Functional Perturbation Theory~\cite{Baroni2001}, as implemented in the \texttt{ph.x} code in the {\tt{Quantum Espresso}} package. In this way, we computed phonon displacements $\textbf{e}^{\nu}(\textbf{Q})$ and energies $\hbar\omega_{\nu,\textbf{Q}}$, together with the Kohn-Sham potential variation $\partial_{j,\alpha}^{\textbf{Q}}V^{KS}$ due to the displacement of the atom $\alpha$ along the direction $j$ in the unit cell.
\\We then computed Kohn-Sham single-particle states on a 24x24x4 $\textbf{k}$-grid, through a non-self-consistent calculation. From this point onwards, crystal symmetries are disabled in our calculations. These wavefunctions are then combined with $\partial_{j,\alpha}^{\textbf{Q}}V^{KS}$ to obtain the electron-phonon coupling matrix elements $g$: this step of the calculation has been carried out with the {\tt{qe2pert.x}} code from the {\tt{Perturbo}} package~\cite{Zhou2021}.
\\We have subsequently exploited the same Kohn-Sham states used for the calculation of $g$ to construct the Bethe-Salpeter Equation (BSE) kernel, as implemented in the {\tt{Yambo}} code~\cite{Marini2009,Sangalli2019}.
The BSE has been solved in the Tamm-Dancoff approximation, using the static approximation for the electron-electron interaction; the electronic screening has been computed within the Random Phase Approximation (RPA), using $120$ bands and a $10$ Ry cutoff for the construction of the RPA electronic polarizability.
The Kohn-Sham single-particle energies used in the independent-particles portion of the BSE kernel have been corrected via a scissor-stretching operator, which correctly reproduces the quasi-particle corrections obtained within the G$_0$W$_0$ approximation of the electron self-energy~\cite{Onida2002}. The highest two (three) valence bands and the lowest two (three) conduction bands have been included in the construction of the BSE kernel for hBN (rBN).
\\Finally, the BSE has been solved at $\textbf{Q} = \Gamma$ without including the long-range term of the exchange kernel, while this contribution has been considered in finite-momentum calculations. We noticed that the inclusion of the macroscopic, longitudinal component of the Coulomb interaction in the calculation of the optical exciton at $\Gamma$ leads to an inverted intensity ratio between the two low-energy peaks
\\The solution of the BSE provides us with both the exciton energies and the envelope functions $A_{\lambda}^{\textbf{Q}}$.
Then, we used {\tt{Python}} post-processing scripts to combine these coefficients $A$ with the electron-phonon matrix elements $g$ to obtain the exciton-phonon couplings $G$, using the definition given in the main text. Note that, as both $A$ and $g$ have been computed starting from the same set of Kohn-Sham states, their phases are coherently defined. 
As a last step, we have used the exciton-phonon couplings to obtain the luminescence spectra, according to Eq.~(1) of the main text: in this equation the summation over $\textbf{Q}$ has been limited to in-plane $\textbf{Q}$-points selected in such a way that the occupation function $N(E_{\lambda}(\textbf{Q}))$ was larger than 10$^{-4}$ for the low-lying excitons $|\lambda,\mathbf{Q}\rangle$. For clarity, we display in Figure \ref{fig:Q_grid_LM} the $\textbf{Q}$-points where finite-momentum excitons have been explicitly evaluated.
%
%
%
%
%
%
\begin{figure*}[ht]
  \centering
    \includegraphics[width=1\textwidth]{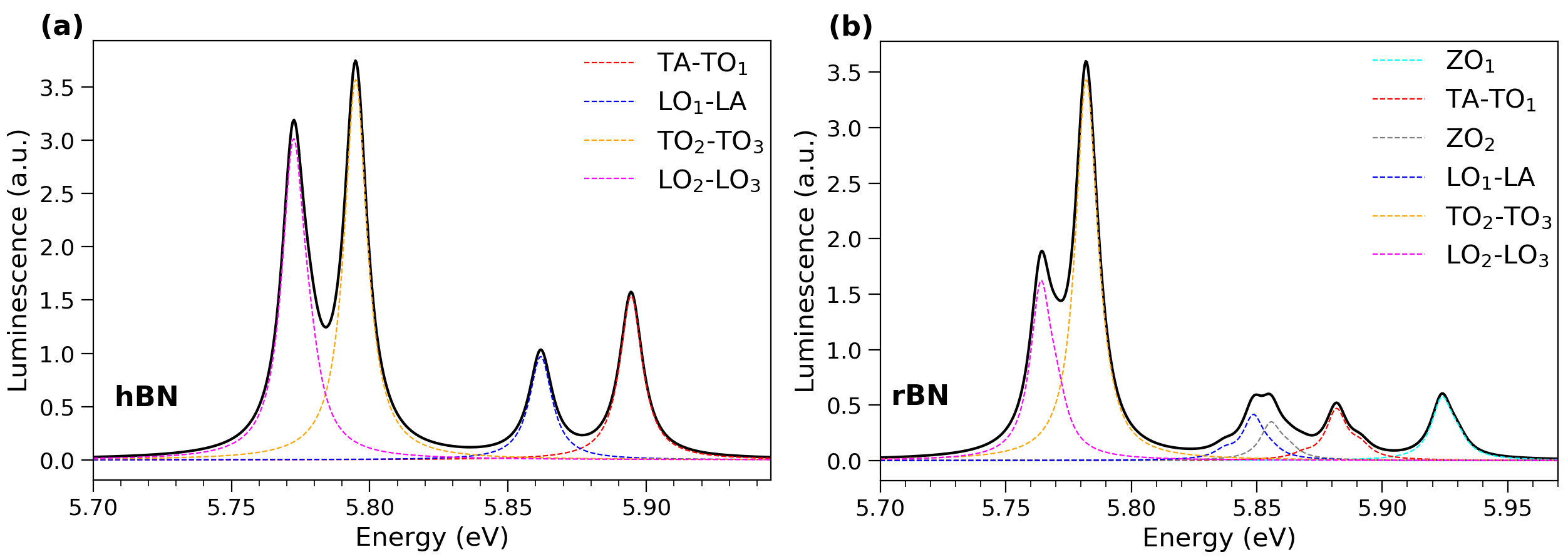}
    \caption{Contribution of different phonon modes in proximity of the $\Omega$ point to the total luminescence (continuous black lines) for hBN (a) and rBN (b). Compare with Fig.~3 of the main text.}
    \label{fig:PH_contrib}
\end{figure*}
\subsection{Exciton dispersion in rBN}
In this section, we present our results for the exciton dispersions. 
The rBN dispersion for the $\mathbf{Q}$-momenta along the $\Gamma$K direction is displayed in Fig. \ref{fig:rBN_exc_disp}(a). 
These results are in very good agreement with the previous calculations by Sponza \textit{et al.}~\cite{Sponza2018} and show that the minimum of the excitonic dispersion occurs close to the $\Omega=(1/6,1/6,0)$ point. (Also for hBN, our dispersion results are in substantial agreement with existing literature\cite{Sponza2018,Schue2019,Chen2019}.)
Figures \ref{fig:rBN_exc_disp}(b) and (c) show the values of the exciton energies in the 2D neighborhoods of the $\Omega$ points where the CL spectrum was computed for both rBN and hBN, respectively (four shells are necessary to converge the CL spectrum for rBN, while three shells are enough for hBN). These plots shows that the true minima of the excitonic dispersion (blue color) do not necessarily lie along a high-symmetry directions and are in fact distributed in a fine structure around $\Omega$. The exact position of the absolute minimum is slightly dependent on the accuracy of the $\mathbf{Q}$-sampling. In general, these plots emphasise that including the effects of the full fine structure around $\Omega$ is important in order to improve the accuracy of the CL spectra with respect to single-$Q$-points methods\cite{Fulvio2019,Cannuccia2019}.
%
%
%

%
%
%
%
%
%
%
\subsection{Phonon-resolved contribution to luminescence}
In this section, we plot the functions 
\begin{equation}
    I_{\nu}(E) = \sum_{\lambda}\sum_{\mathbf{Q}}N(E_{\lambda}(\mathbf{Q}))\Gamma_{\lambda}^{\nu,\mathbf{Q}}(E)
\end{equation}
which correspond to the contributions of different phonon branches to the total luminescence spectra.
These functions are shown in Fig.\ref{fig:PH_contrib} (the contributions of phonon modes which are very close in energy in proximity of the $\Omega$ point are summed). The phonon branches are labelled using the same notation presented in the main text.
Note in particular the important role of out-of-plane (Z) phonon modes in the rBN case.
We also point out that a theoretical luminescence calculation not including exciton-phonon interactions would completely miss this multi-peak fine structure. In fact, it would just give a single dominant peak at the energy of the lowest-lying bright exciton, which lies around 250 meV above the phonon-assisted peaks and is therefore not observed in experiment.
\FloatBarrier
%